\documentclass[useAMS]{mn2e}
\usepackage{psfig}
 \usepackage{url}
\usepackage{times}
\def\spose#1{\hbox to 0pt{#1\hss}}
\newcommand\lsim{\mathrel{\spose{\lower 3pt\hbox{$\mathchar"218$}}
     \raise 2.0pt\hbox{$\mathchar"13C$}}}
\newcommand\gsim{\mathrel{\spose{\lower 3pt\hbox{$\mathchar"218$}}
     \raise 2.0pt\hbox{$\mathchar"13E$}}}
\def\sc{Schwarzschild}
\def\ltsima{$\; \buildrel < \over \sim \;$}
\def\lsim{\lower.5ex\hbox{\ltsima}}
\def\gtsima{$\; \buildrel > \over \sim \;$}
\def\gsim{\lower.5ex\hbox{\gtsima}}

\title[SDSS J114657.79+403708.6: a blazar at $z$=5.0]
{SDSS J114657.79+403708.6: the third most distant blazar at $z$=5.0}
\author[G. Ghisellini et al.]
{G. Ghisellini$^1$, \thanks{E--mail: gabriele.ghisellini@brera.inaf.it}
T.  Sbarrato$^{1,2,3}$, G. Tagliaferri$^1$, L. Foschini$^1$, F. Tavecchio$^1$, G. Ghirlanda$^1$,  
\newauthor{
V. Braito$^1$, N. Gehrels$^4$ }  \\
$^1$ INAF -- Osservatorio Astronomico di Brera, via E. Bianchi 46, I--23807 Merate, Italy \\
$^2$ Univ. dell'Insubria, Dipartimento di Fisica e Matematica, Via Valleggio 11, I--22100 Como, Italy \\
$^3$ ESO--European Southern Observatory, Karl--Schwarzschild-Strasse 2, 8578 Garching bei M\"unchen, Germany\\
$^4$ NASA--Goddard Space Flight Center, Greenbelt, Maryland 2077, USA
}

\begin{document}

\pagerange{\pageref{firstpage}--\pageref{lastpage}} \pubyear{2012}

\maketitle
\label{firstpage}

\begin{abstract}
The radio--loud quasar SDSS J114657.79+403708.6
at a redshift $z$=5.0 is one of the most distant radio--loud objects.
The IR--optical luminosity and spectrum suggest that its black hole has a very large
mass: $M=(5\pm 1) \times 10^9 M_\odot$. 
The radio--loudness (ratio of the radio to optical flux) 
of the source is large (around 100), suggesting that the source is viewed
at small angles from the jet axis, and could be a blazar.
The X--ray observations fully confirm this hypothesis, due to the
high level and hardness of the flux.
This makes SDSS J114657.79+403708.6 the third most distant blazar known,
after Q0906+693 ($z=5.47$) and B2 1023+25 ($z=5.3$).
Among those, SDSS J114657.79+403708.6 has the largest black hole mass,
setting interesting constraints on the mass function of heavy ($>10^9 M_\odot$)
black holes at high redshifts.
\end{abstract}



\section{Introduction}
\label{sec-intro}

Since the radiation produced by relativistic jets is strongly
boosted along the jet direction, objects whose jet is pointing at us
are very bright, and can be seen up to high redshifts.
The black hole at the center of the powerhouse of these
sources can be very massive, sometimes exceeding $M=10^{10}M_\odot$ 
(Ghisellini et al. 2009; 2010a). 
Therefore the hunt for high redshift blazars is an important field of research,
allowing the census of heavy black holes in the early Universe.
This can confirm or challenge existing theories of black hole formation, 
and even more so if we associate the presence of relativistic jets with
a large black hole spin. 
In this case, in fact, the efficiency of accretion $\eta$ (the fraction of accreted mass
transformed into radiation) is higher than for a non--rotating black hole,
reaching a value of $\eta=0.3$ for maximally rotating {\it accreting} black holes 
(see Thorne 1974).
This implies that the Eddington luminosity is reached with a smaller accretion
rate than for a non--rotating black hole.
If the system is Eddington limited, this in turn implies a slower
black hole growth. 
If a black hole seed starts to accrete at a redshift $z=20$ at the Eddington rate
maintaining a large spin, it can reach a billion solar masses only at $z<4$ even for a seed
mass as large as $10^6 M_\odot$ (see e.g. Ghisellini et al. 2013).
The very fact of the existence of radio--loud sources with $M>10^9 M_\odot$
at $z>4$ is thus a problem.

These issues motivate our search of high redshift blazars, keeping in mind that for
each blazar (i.e. viewing angle smaller than $1/\Gamma$, where $\Gamma$ is the jet bulk Lorentz factor) 
there must be other $2\Gamma^2$ sources whose jets are pointing in other directions.

Up to now there are two blazars known at $z>5$:
Q0906+6930 ($z=5.47$, Romani et al. 2004; Romani 2006), 
and B2 1023+25 ($z=5.3$, Sbarrato et al. 2012, 2013).
The spectral energy distribution (SED) of these sources
reveals both the thermal (i.e. strong optical emission lines and continuum) and the
boosted non--thermal components.
As in the majority of very powerful and high--$z$ blazars, the thermal disk 
emission becomes visible since it stands between the two non--thermal humps (the synchrotron
one peaking in the sub-mm, and the high energy in the $\sim$MeV band; Ghisellini et al. 2010a; 2010b).
In these sources the  X--ray spectrum is hard 
[i.e. $\alpha_X \sim 0.5$, assuming $F(\nu) \propto \nu^{-\alpha_{x}}$], 
and this, together with a relatively strong X--ray to optical flux ratio, can be taken as a signature
of the blazar nature of the source.

In this letter we suggest that SDSS J114657.79+403708.6, 
a radio--loud AGN at $z=5.005$,
is a blazar, i.e. the viewing angle is smaller than $1/\Gamma$.
Evidences for its blazar nature include its relatively large radio--loudness 
and its very large X--ray luminosity and hard X--ray spectrum, as measured
by a pointed {\it Swift} observation.
Infrared data collected by the WISE satellite (Wright et al. 2010), together with the
Sloan Digital Sky Survey (SDSS; York et al. 2000) spectrum  allowed to constrain the
properties of the thermal emission.

In this work, we adopt a flat cosmology with $H_0=70$ km s$^{-1}$ Mpc$^{-1}$ and
$\Omega_{\rm M}=0.3$.

\section{SDSS J114657.79+403708.6 as a blazar candidate}
\label{sec-sample}

SDSS J114657.79+403708.6 (SDSS 1146+403 hereafter) 
belongs to the SDSS DR7 Quasar catalog 
(Schneider et al. 2010), that have been analysed by Shen et al. (2011). 
This sample contains $\sim$105,000 quasars.
The area of the sky surveyed by the SDSS has been almost completely sampled
by the the FIRST (Faint Images of the Radio Sky at Twenty-cm; Becker, White \& Helfand, 1995) 
survey, with a flux limit of 1 mJy at 1.4 GHz.
The sky area covered by both surveys is $\sim$8,800 square degrees.
SDSS 1146+403 is detected in the radio with a flux of $\sim$13 mJy at 1.4 GHz,
while VLBI observations yielded a flux of $15.5\pm0.8$ and $8.6\pm0.4$ mJy at 1.6 and 5 GHz,
respectively (Frey et al. 2010).
These fluxes correspond to a $\nu L(\nu)$ radio luminosity of $\sim 10^{44}$ erg s$^{-1}$.
Comparing the radio and the optical flux we obtain a radio loudness of $\sim$100,
calculated by assuming a flat radio spectrum (i.e. $F(\nu)\propto \nu^0$) and
calculating the rest frame 2500 \AA\ flux (which is not observed) by extrapolating 
the continuum to $\lambda=1350\rm\AA$.

SDSS 1146+403 is included in the 
AllWISE Source Catalog\footnote{
Data retrieved from \url{http://irsa.ipac.caltech.edu/}}, 
with clear detections in the two bands at lower wavelengths of the instrument, 
i.e.\ $\lambda=3.4\mu$m and $\lambda=4.6\mu$m.
The source is not detected by the Large Area Telescope (LAT) onboard the {\it Fermi}
satellite.

\subsection{{\it Swift} observations}

The large radio loudness of SDSS 1146+403 suggests a small viewing angle, and
to confirm its blazar nature we observed the source with the {\it Swift}
satellite (Gehrels et al. 2004).
In fact the X--ray spectrum of FSRQs beamed towards us is particularly 
bright and hard, with an energy spectral index $\alpha_x \sim 0.5$ [$F(\nu) \propto \nu^{-\alpha_x}$]
(see e.g. Ghisellini et al. 2010b, Wu et al. 2013).

The observations were performed between Jan. 16 and Jan 30 2013
(ObsIDs: 00049384001, 00049384002, 00049384003, 00049384004, 00049384005, 00049384007, 00049384008, 00049384009.). 
The ObsID 00049384001 e 00049384009 did not generate spectra nor light--curves.

Data of the X--ray Telescope (XRT, Burrows et al. 2005) and the UltraViolet Optical Telescope 
(UVOT, Roming et al. 2005) were downloaded from HEASARC public archive, 
processed with the specific {\it Swift} software included in the package 
{\tt HEASoft v. 6.15} and analysed. 
The calibration database was updated on 12, 2013.
We did not consider the data of the Burst Alert Telescope (BAT, Barthelmy et al. 2005), 
given the weak X--ray flux.

The total exposure on the XRT was $\sim$40.8 ks.
The mean count rate was  $(1.74\pm 0.21)\times 10^{-3}$, resulting in 71 total counts. 
Given the low statistics, the fit with a power law model with Galactic absorption 
($N_{\rm H}=1.65\times 10^{20}$~cm$^{-2}$, Kalberla et al. 2005) 
was done by using the likelihood (Cash 1979). 
The output parameters of the model were a photon spectral index
$\Gamma_x =(\alpha_x+1)= 1.5\pm 0.3$ 
and an integrated observed flux 
$F_{0.3-10\,\rm keV}=(1. \pm 0.12)\times 10^{-13}$~erg~cm$^{-2}$~s$^{-1}$. 
The value of the likelihood was 54.93 for 63 dof.
The X--ray data displayed in the SED (Fig. \ref{sed}) has been rebinned 
to have $3\sigma$ in each bin. 

UVOT observed the source only in the $v$ filter. 
The total was 61.2 ks.
The source was not detected, and can derive a 3$\sigma$ upper limit $v>22.46$ mag.
We expect, along the line of sight, some absorption due to the intervening matter.
A rough estimate of the optical depth for the UVOT $v$ filter for $z=5$ gives $\tau\sim 2$
(F. Haardt, priv. comm., see also Ghisellini et al. 2010a).
Using this value of $\tau$, the unabsorbed upper limit on the flux in the $v$ filter
becomes $F<0.03$ mJy.

\subsection{Estimate of the disc luminosity}
\label{Ld}

We can infer the disc luminosity $L_{\rm d}$ in two ways.
The first is by assuming that the IR--optical flux is due to accretion.
The {\it WISE} IR data together with the optical spectrum redward of the 
Ly--$\alpha$ SDSS 1146+403 provide a  set of IR--optical data that show
a rising (in $\nu L_\nu$) slope and a peak around $\sim 2\times 10^{15}$ Hz (rest frame),
below the absorption caused by intervening clouds (see Fig. \ref{mass}).
The $\nu L_\nu$ luminosity at this peak is $\sim 5\times 10^{46}$ erg s$^{-1}$, 
corresponding to a integrated IR--optical luminosity of $10^{47}$ erg s$^{-1}$. 
We associate this emission to the accretion disk.
We can then immediately find the accretion rate assuming the accretion efficiency.

The second method relies on the presence of broad emission lines,
that re--emit a fraction $C$ (i.e. their covering factor) of the
ionizing luminosity, assuming that the latter is provided by
the accretion disc.
The average value of $C$ is around 0.1 (Baldwin \& Netzer, 1978; Smith et al., 1981),
with a rather large dispersion.
This method, discussed in detail in Calderone et al. (2013), is based 
on templates of line to bolometric luminosities ratios, as the ones presented
in Francis et al. (1991) or Vanden Berk et al. (2001).
For instance in Francis et al. (1991) the ratio of the total luminosity
of the broad line region (BLR hereafter) to the Ly$\alpha$ line luminosity is 5.5,
while in van der Berk (2001) is 2.7.
The SDSS optical spectrum of 1146+403 shows a prominent Ly$\alpha$+NV line 
complex, which is absorbed in its blue part. 
Doubling the luminosity of only the red part, we obtain $L_{Ly\alpha}\sim 3\times 10^{45}$
erg s$^{-1}$. 
This corresponds to $L_{\rm BLR}$=(0.8--1.7)$\times 10^{46}$ erg s$^{-1}$.
If $C=0.1$, the disk luminosity is then 10 times greater.
Within the relevant uncertainties, the agreement between the two methods
is rather satisfactory.

\subsection{Estimate of the black hole mass}
\label{mass}

A robust lower limit to the black hole mass of SDSS 1146+403
is set by assuming that its disc emits at the Eddington luminosity.
This implies $M>10^9 M_\odot$.

The commonly used virial method, based on the relation between the ionizing
luminosity and the distance of the BLR, and on the assumption that
the velocity of the clouds is virialized and it is linked to gravity 
(e.g. Wandel 1997; Peterson et al. 2004)
is not (yet) applicable in this case, because we do not have the appropriate
scaling relations for the Ly$\alpha$ line
(which in addition has an uncertain FWHM, since its blue wing is absorbed)
and we do not have (yet) an IR spectrum of the source that can reveal
the CIV and the MgII broad lines.

We then apply the method of fitting a disk accretion model spectrum to the data.
We assume a simple Shakura \& Sunyaev disk (1973) model, which depends only on the
black hole mass $M_{\rm BH}$ and on the accretion rate $\dot{M}$.
The latter is traced by the total disc luminosity $L_{\rm d}=\eta\dot{M}c^2$.
For simplicity, we assume that the black hole is non rotating and
set $\eta=0.08$ and the last stable orbit at 3 \sc\ radii.
If we measure $L_{\rm d}$ (hence $\dot M$), we are left with $M_{\rm BH}$
as the only remaining free parameter. For a given $\dot M$, the black hole mass
determines the peak frequency of the disc emission.
A larger mass implies a larger disc surface and hence a lower temperature 
emitting a given luminosity. 
Therefore, for a fixed $L_{\rm d}$, a larger $M_{\rm BH}$ shifts the peak 
to lower frequencies. 
Then the best agreement with the data fixes $M_{\rm BH}$ (e.g. Calderone et al. 2013).
Using a larger efficiency would enhance the emission at high (rest frame UV) frequencies.
Assuming that the disk emits the same  $L_{\rm d}$, this implies a smaller flux 
below the disk emission peak.
To match the optical--IR emission one would then require a larger black hole mass
(see the discussion on this issue in Calderone et al. 2013).

Fig. \ref{mass} shows the data in the IR--optical band (as labelled) and three 
disc emission spectra calculated assuming the same $L_{\rm d}$ and 
three different mass values:
$M_{\rm BH}=6\times10^9M_\odot$ (red line), 
$5\times10^9M_\odot$ (blue) and $4\times10^9M_\odot$ (green).
These values are indicative of the uncertainties on the black hole mass,
which is therefore $(5\pm1)\times 10^9 M_\odot$.

\begin{figure}
\hskip -0.3cm
\psfig{file=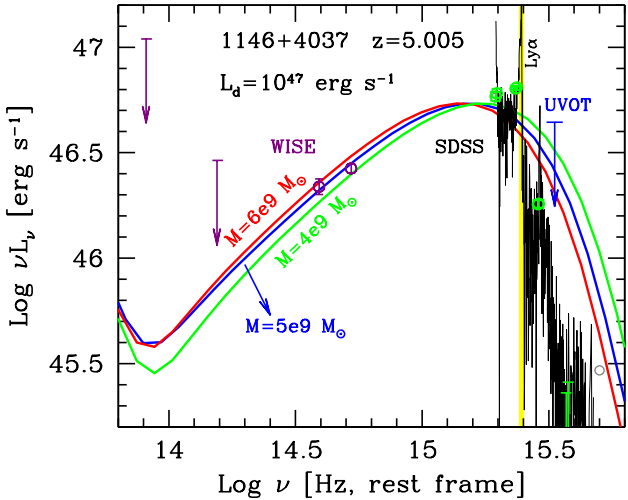,height=9.2cm,width=9.2cm}
\vskip -0.7 cm
\caption{
Optical--UV SED of SDSS 1146+403 in the rest frame, together with 
models of standard accretion disc emission.
Optical and UV data, including the SDSS spectrum, have been corrected
for the extinction in our Galaxy.
Data from WISE 
(AllWISE Source Catalog)
and GROND, and the SDSS spectrum are labelled.
Green points are archival data taken from ASDC SED builder.
The two optical photometric points receive some contribution
from the emission line flux, besides the continuum.
We show the spectrum of three accretion disc models
with the same luminosity and different $M_{\rm BH}$:
$M_{\rm BH}/M_\odot=6\times 10^9$ (red), $5\times 10^9$ (blue) and 
$4\times 10^9$ (green).
Note that outside this range of masses, the model cannot fit 
satisfactory the data.
}
\label{mass}
\end{figure}

\begin{table*} 
\centering
\begin{tabular}{lllllllllllllllll}
\hline
\hline
Name   &$z$ &$R_{\rm diss}$ &$M$ &$R_{\rm BLR}$ &$P^\prime_{\rm i}$ &$L_{\rm d}$ &$L_{\rm d}/L_{\rm Edd}$ 
&$B$ &$\Gamma$ &$\theta_{\rm v}$
    &$\gamma_{\rm b}$ &$\gamma_{\rm max}$  &$P_{\rm r}$ &$P_{\rm B}$ &$P_{\rm e}$ &$P_{\rm p}$ \\
~[1]      &[2] &[3] &[4] &[5] &[6] &[7] &[8] &[9] &[10] &[11]  &[12] &[13]  &[14] &[15] &[16] &[17]\\
\hline   
1146+430   &5.005 &900  &5e9   &1006  &7e--3 &100  &0.15  &1.4 &13  &3  &230 &3e3   &3.7  &10.3 &0.05 &15.1 \\ 
1146+430   &5.005 &900  &5e9   &1006  &0.8   &100  &0.15  &1.5 &6   &12 &50  &3e3   &75.9 &2.3  &4.6  &741  \\
\hline
0906+693   &5.47  &630  &3e9   &822   &0.02  &67.5 &0.17  &1.8 &13  &3  &100 &3e3   &10.4  &8.2  &0.2  &58  \\   
1023+25    &5.3   &504  &2.8e9 &920   &0.01  &90   &0.25  &2.3 &13  &3  &70  &4e3   &5    &8.5  &0.14 &40.7  \\    
\hline
\hline 
\end{tabular}
\vskip 0.4 true cm
\caption{List of parameters adopted for or derived from the model for the ED of SDSS 1146+403, 
compared with the set of parameters used for the other two blazars at $z>5$.
Col. [1]: name;
Col. [2]: redshift;
Col. [3]: dissipation radius in units of $10^{15}$ cm;
Col. [4]: black hole mass in solar masses;
Col. [5]: size of the BLR in units of $10^{15}$ cm;
Col. [6]: power injected in the blob calculated in the comoving frame, in units of $10^{45}$ erg s$^{-1}$; 
Col. [7]: accretion disk luminosity in units of $10^{45}$ erg s$^{-1}$  
Col. [8]: $L_{\rm d}$ in units of $L_{\rm Edd}$;
Col. [9]: magnetic field in Gauss;
Col. [10]: bulk Lorentz factor at $R_{\rm diss}$;
Col. [11]: viewing angle in degrees;
Col. [12] and [13]: break and maximum random Lorentz factors of the injected electrons;
Col. [14]: power spent by the jet to produce the non--thermal beamed radiation in units of $10^{45}$ erg s$^{-1}$;
Col. [15]: jet Poynting flux in units of $10^{45}$ erg s$^{-1}$; 
Col. [16]: power in bulk motion of emitting electrons, in units of $10^{45}$ erg s$^{-1}$.
Col. [17]: power in bulk motion of cold protons, assuming one proton per emitting electron in units of $10^{45}$ erg s$^{-1}$.
The total X--ray corona luminosity is assumed to be in the range 10--30 per cent of $L_{\rm d}$.
Its spectral shape is assumed to be always $\propto \nu^{-1} \exp(-h\nu/150~{\rm keV})$.
}
\label{para}
\end{table*}

\begin{figure}
\hskip -0.3cm
\psfig{file=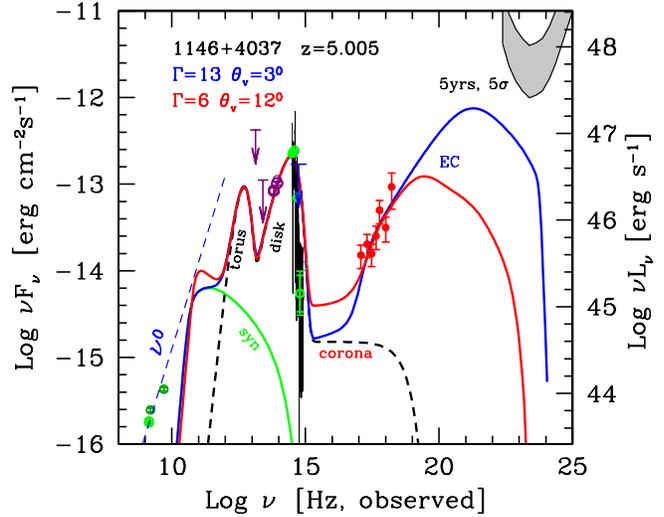,height=9.cm,width=9.cm}
\vskip -0.5 cm
\caption{
SED of SDSS 1146+403 together with the adopted 
models, as labelled. 
Data from WISE, GROND, {\it Swift}/XRT and {\it Fermi}/LAT
are labelled.
Green points are archival data taken from ASDC SED builder.
The solid green line is the synchrotron component of the model,
the dashed black line is the torus+disc+X--ray corona 
component.
The high energy emission is dominated by the External Compton (EC) component.
The lower bound of the grey stripe correspond to 
the LAT upper limits for 5 years, 5$\sigma$.
} 
\label{sed}
\end{figure}

\begin{figure}
\hskip -0.3cm
\psfig{file=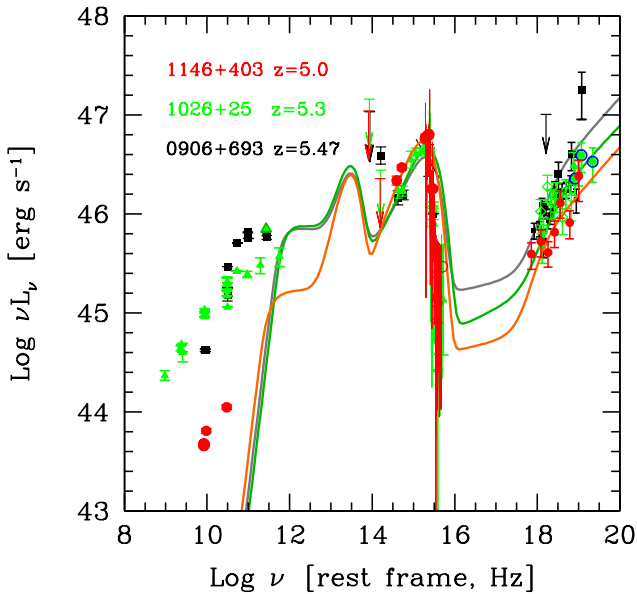,height=9cm,width=9cm}
\vskip -0.5 cm
\caption{
Comparison between the SED of SDSS 1146+403 (red symbols) with B2 1023+25 (green symbols) 
and  Q0906+6930 (black symbols). 
The solid lines corresponds to the models whose parameters are listed in Tab. \ref{para}
(for 1146+403 we show the model with $\Gamma=13$ and $\theta_{\rm V}=3^\circ$).
The three sources are very similar in the IR--optical and X--rays, but
SDSS 1146+403 is a factor $\sim$20 less luminous in the radio.
The strong X--ray luminosity (with respect to the optical) and the hard X--ray spectra 
flag the presence of jet radiation beamed in the observer's direction.
The difference in the radio band could be due to a slight bend of the jet 
between the regions emitting X--rays (inner jet) and the radio (more external jet).
} 
\label{comparison}
\end{figure}

\section{Overall spectral energy distribution}

Fig. \ref{sed} shows the overall SED of SDSS 1146+403 from radio to $\gamma$--rays.
The optical data and the SDSS spectrum have been corrected 
for galactic extinction.
The dashed line in the radio domain indicates a flat radio spectrum $F_\nu \propto \nu^0$),
while the grey hatched area corresponds to the limiting sensitivity of {\it Fermi}/LAT.
The lower boundary is the 5$\sigma$ flux limit for 5 years of {\it Fermi} operations.
The same figure shows the result of a fitting model, described in detail in Ghisellini \& Tavecchio (2009).
It is a one--zone, leptonic model, in which relativistic electrons
emit by the synchrotron and Inverse Compton processes.
The electron distribution is derived through a continuity equation,
assuming continuos injection, radiative cooling, possible pair production
and pair emission, and is calculated at a time $R/c$ after the start of the injection, 
where $R$ is the size of the source, located at a distance $R_{\rm diss}$ from the black hole.
The jet is viewed at an angle $\theta_{\rm v}$ from the jet axis.
The accretion disk component is accounted for, as well the
infrared emission reprocessed by a dusty torus and the X--ray 
emission produced by a hot thermal corona sandwiching the accretion disc.
We present two models.
The first ($\Gamma=13$; $\theta_{\rm v}=3^\circ$)
is the best representation of the data assuming a set of parameters 
very similar to other powerful blazars (Ghisellini et al., 2010a; 2010b), while the second assumes 
$\Gamma=6$ and $\theta_{\rm v}=12^\circ$.
This latter choice corresponds to the model with the maximum viewing angle compatible with the data
and a still reasonable bulk Lorentz factor.
Tab. \ref{para} reports the relevant parameters of the two models of SDSS 1146+403,
together with the parameters for the other two blazars at $z>5$.

Both models fit the X--radio data and are very similar in the radio band.
Having the same black hole mass and accretion rate, they have
the same shape and flux in the IR--optical range. 
However, they correspond  to jets with a very different intrinsic power. 
The small $\Gamma$, relatively large $\theta_{\rm v}$ solution implies a much
less beamed emission, and therefore demands a greater intrinsic power.
This leads us to prefer the more ``economic" solution.
Note that the models differ in the hard X--ray range, where the source could be detected by 
the {\it NuSTAR} satellite (sensitive up to $\sim$80 keV; Harrison et al. 2013).

\subsection{Comparison with Q0906+6930 and B2 1023+25}

Fig. \ref{comparison} shows how the SED of 1146+403 compares with the SED
(in $\nu L_\nu$ vs rest frame $\nu$)
of Q0906+693 ($z$=5.47) and B2 1023+25 ($z$=5.3).
As can be seen, the three SEDs are very similar.
All three sources have approximately the same IR--optical 
spectrum, and also the X--ray spectra are remarkably similar.
The main difference is in the radio band, with SDSS 1146+403 
being under--luminous by a factor $\sim$20 with respect to the other two blazars.

The X--ray emission in blazars is usually produced in a rather compact region 
of the jet, possibly within the BLR. 
This is the most efficient way to produce high energy photons
through inverse Compton scattering, due to the presence of emission line
photons, besides the synchrotron photons produced internally to the jet.
Furthermore, a compact region can account for the short time variability
often shown by blazars. 
On the other hand, if the X--ray flux is variable, the approximate equal
X--ray flux detected in the three sources must be a coincidence.
The lack of strong variability could be explained, in the adopted model,
by the fact that the X--rays are produced by low energy electrons
(with random Lorentz factors $\gamma\lsim$2--5) scattering emission line photons:
since the cooling time $\propto 1/\gamma$, these electrons 
may have not time to cool in one dynamical time $R/c$. 
Consequently, the X--ray flux they produce is less variable than at higher 
($\gamma$--ray) energies, characterized by a much shorter cooling time.

The size of the region emitting the radio flux at $\sim$ GHz frequencies
must be much larger (and therefore more external) 
than the X--ray region, since otherwise the flux would be self--absorbed. 
This is the reason why our one--zone model cannot reproduce the radio below 
a several tens of GHz.
The smaller radio flux of SDSS 1146+403 could then be due to a jet bending
between the X--ray and the radio regions.
Assuming a constant $\Gamma=13$, the radio deficit (factor $\sim$20) 
can be explained by a change in viewing angle from $\theta_{\rm v}=3^\circ$ 
(corresponding to the X--ray production region of the jet) 
to $\theta_{\rm v}=6^\circ$ (corresponding to the jet region producing the radio).

Alternatively, the jet could decelerate between the two regions maintaining the same viewing angle:
in this case $\Gamma$ must decrease by a factor $20^{1/4}\sim 2.1$.
However, in this case, the deceleration should correspond to a relevant dissipation,
in turn corresponding to some emission, which we do not see.
It would be interesting to observe the source at high radio 
frequencies (100--300 GHz, rest frame), to see if the radio deficit 
remains the same or becomes smaller.

\section{Discussion and conclusions}
\label{sec-discussion}

In this letter we propose that the radio--loud, high redshift quasar SDSS 1146+403
is a blazar.
If so, it is the third most distant blazar known up to now, with redshift 
$z=5.005$, that corresponds to a cosmic age 1.1 billion years.
Despite this young age, the black hole of SDSS 1146+403 managed to grow
to 5 billion solar masses.

Both its thermal and non--thermal components are very luminous. 
In agreement with the blazar sequence (Fossati et al. 1998) the two broad 
non--thermal humps peak at small frequency.
In particular, the hard X--ray spectrum and the upper limit in the $\gamma$--ray band 
constrain the high energy component to peak in the MeV region of the spectrum.
Therefore this source, along with the similar other powerful blazars,
should have a relatively large hard X--ray flux and would have been an
ideal target for hard X--ray instruments such as
the focussing hard X--ray telescope {\it NuStar} (Harrison et al., 2013). 
The great sensitivity over its energy range [5--80 keV] would enable it to 
detect the hard X--ray spectrum of this source, even if it cannot 
directly observe the peak of the high energy hump.

\vspace{0.3 cm}
It is possible to roughly estimate the number of these objects and their
spatial density.

The comoving density of heavy black holes at high redshifts
of radio--loud sources has been studied by Volonteri et al. (2011),
based on the 3 years BAT catalog and the blazar luminosity function,
in hard X--rays, derived by Ajello et al. (2009), and modified
(beyond $z=4.3$) by Ghisellini et al. (2010a).
In the latter paper the observational constrain on the blazar density
with black holes heavier than $10^9 M_\odot$ in the redshift bin
$5<z<6$ was based on the detection of only one object: Q0906+6930.
Assuming it was the only blazar in the entire sky in this
redshift bin, Ghisellini et al. (2010a) derived a comoving density of 
$2.63\times 10^{-3}$ Gpc$^{-3}$ of blazars hosting an heavy black hole
in the $5<z<6$ bin (see Fig. 15 in that paper).

SDSS 1146+403 was selected in the SDSS catalog covered by FIRST observations,
and the common area of the sky of these two surveys is 8770 square degrees.
It is the second source that can be classified as a blazars in this
SDSS+FIRST survey (together with B2 1023+25).
The comoving volume in the redshift bin $5<z<6$ is 380 Gpc$^3$.
Therefore the number density of blazars in this redshift
bin is $N_{BL}= 2\times (40,000/8,770)/ 380 = 2.4\times 10^{-2}$ Gpc$^{-3}$.
Both B2 1023+25 and SDSS 1146+403 have black holes with $M>10^9 M_\odot$.
To find out the density of black holes in jetted sources 
heavier than one billion solar masses,
we should consider that for each jet observed within a viewing angle $\theta_{\rm v}=1/\Gamma$,
there exist another $2\Gamma^2$ sources pointing in other directions.
So the number density of black holes with $5<z<6$, with a mass 
exceeding $10^9M_\odot$, is $0.024 \times 2 \times 169 (\Gamma/13)^2=8.1$ Gpc$^{-3}$.
Multiplying by the comoving volume gives 
$\sim$3,000 heavy black holes in jetted sources only. 

As mentioned in the introduction, the finding of heavy and early black holes 
in sources with jets can severely challenge our understanding of black hole
growth, especially if we associate the presence of the jet with a 
rapidly spinning black hole. 
A Kerr black hole is in fact more efficient to transform
gravitational energy into radiation than a non spinning (radio--quiet)
black hole.
This lends support to the possibility of super--Eddington accretion
(Volonteri \& Silk 2014), and/or to the possibility that part of the
gravitational energy of the accreting matter is not used to heat the
disk, but to amplify the magnetic fields necessary to extract the rotational
energy of the black hole (see e.g. Jolley \& Kunzic 2008; Shankar et al. 2008, Ghisellini et al. 2013 ).

 
\section*{Acknowledgements}
We thank the anonymous referee for useful comments.
This publication makes use of data products from the Wide--field
Infrared Survey Explorer, which is a joint project of the University
of California, Los Angeles, and the Jet Propulsion
Laboratory/California Institute of Technology, funded by NASA.
Part of this work is based on archival data and on--line service provided by the ASI 
Science Data Center (ASDC).



\label{lastpage}
\end{document}